\documentclass[aps,prl,twocolumn,showpacs,superscriptaddress,groupedaddress]{revtex4}  % for review and submission
\usepackage{graphicx}  % needed for figures
\usepackage{amssymb}   % for math
\usepackage{amsmath}

\usepackage{hyperref}
\usepackage{xcolor}

\usepackage{endnotes}
\let\footnote=\endnote

\def\D{{\cal D}}
\def\g{\overline {\cal G}}

\def\ep{\epsilon}
\def\zo{\overline{z}_1}
\def\zt{\overline{z}_2}
\def\C{\overline{C}}

\begin{document}

\begin{flushleft} 
\mbox{HRI-RECAPP-2014-008}
\end{flushleft}

\title{Rapidity Distributions in Drell-Yan and Higgs Productions 
at Threshold to Third Order in QCD}

\author{Taushif Ahmed\footnote{taushif@hri.res.in}}
\affiliation{Regional Centre for Accelerator-based Particle Physics,
Harish-Chandra Research Institute,  
Allahabad, India}

\author{M. K. Mandal\footnote{mandal@hri.res.in}}
\affiliation{Regional Centre for Accelerator-based Particle Physics,
Harish-Chandra Research Institute, 
Allahabad, India}

\author{Narayan Rana\footnote{narayan@hri.res.in}}
\affiliation{Regional Centre for Accelerator-based Particle Physics,
Harish-Chandra Research Institute, 
Allahabad, India}

\author{V. Ravindran\footnote{ravindra@imsc.res.in}}
\affiliation{The Institute of Mathematical Sciences, 
Chennai, India }

\date{\today}

\begin{abstract}
We present the threshold N$^3$LO perturbative QCD corrections to the rapidity 
distributions of dileptons in the Drell-Yan process and Higgs boson in gluon fusion.  
Sudakov resummation of QCD amplitudes, renormalization group invariance,  
and the mass factorization theorem provide useful guidelines
to obtain them in an elegant manner.   
We use various state of the art three loop results that have been recently 
available to obtain these distributions.  
For the Higgs boson, we demonstrate numerically the importance of these corrections at the LHC. 
\end{abstract}

\maketitle
Drell-Yan (DY) production \cite{Drell:1970wh} of a pair of leptons at the Large
Hadron Collider (LHC) is one of the cleanest processes that can
be studied not only to test the standard model (SM) to an 
unprecedented accuracy but also to probe physics beyond
the SM (BSM) scenarios in a very clear environment. Rapidity
distributions of $Z$ boson \cite{Affolder:2000rx} and charge 
asymmetries of leptons in $W$ boson decays \cite{Abe:1998rv} constrain various
parton densities and, in addition, possible excess events can provide
hints to BSM physics, namely R-parity violating supersymmetric models,
models with $Z'$ or with contact interactions and large
extra-dimension models.
One of the production mechanisms responsible for
discovering the Higgs boson of the SM at the LHC \cite{AtlasCMS} is
the gluon-gluon fusion through top quark
loop. Being a dominant one, it will continue to play a major role in studying the properties of the Higgs boson and its coupling to other SM particles. 
Distributions of transverse momentum and rapidity of the Higgs boson are going to be
very useful tools to achieve this task. Like the inclusive
rates \cite{inclratesDY, inclratesH}, the rapidity distribution 
of dileptons in DY production 
and of the Higgs boson in gluon-gluon fusion are also known to next to next to leading order
(NNLO) level in perturbative QCD due to seminal works
by Anastasiou \textit{et al.}~\cite{Anastasiou:2003yy}.
The quark and gluon form factors \cite{3lffmoch, Baikov:2009bg, Gehrmann:2010ue},
the mass factorization kernels \cite{Moch:2004pa}, and the renormalization 
constant \cite{renorm} for the effective operator describing the coupling of the
Higgs boson with the SM fields in the infinite top quark mass limit up to three loop level in dimensional regularization
with space-time dimensions $n = 4 + \epsilon$ were found to be useful to obtain
the next to next to next to leading order (N$^3$LO) threshold effects \cite{n3losv} to the
inclusive Higgs boson and DY productions at the LHC, excluding $\delta(1-z)$ terms,
where the scaling parameter is $z=m_{l^+l^-}^2/\hat s$ for the DY process and 
$z=m_H^2/\hat s$ for the Higgs boson. 
Here, $m_{l^+l^-}$, $m_H$ and $\hat s$ are the
invariant mass of the dileptons, the mass of the Higgs boson,
and center of mass energy of the partonic reaction responsible for the production mechanism, 
respectively.  
Recently, Anastasiou \textit{et al.}~\cite{Anastasiou:2014vaa} made an important contribution in computing 
the total rate for the Higgs boson production at N$^3$LO resulting from the threshold
region including the $\delta(1-z)$ term.  Their result, along with
three loop quark form factors and mass factorization kernels, was used to 
compute the DY cross section at N$^3$LO at threshold in \cite{Ahmed:2014cla}.

In this Letter, we will apply the formalism developed in \cite{Ravindran:2006bu} to obtain 
rapidity distributions of the dilepton pair and of the Higgs boson at N$^3$LO in the threshold
region using the available information that led to the computation 
of the N$^3$LO threshold corrections to the inclusive Higgs boson and DY productions.

The rapidity distribution can be written as
\begin{eqnarray}\label{sighad}
{d \sigma^I \over dy } =
\sigma^I_{\rm Born}(x_1^0,x_2^0,q^2) W^I(x_1^0,x_2^0,q^2) ,\quad
I=q, g,
\end{eqnarray}
normalized by
$W^I_{\rm Born}(x_1^0,x_2^0,q^2)=\delta(1-x_1^0)\delta(1-x_2^0)$.
Rapidity~~
$y={1 \over 2} \log(p_2.q/p_1.q)={1 \over 2 } \log\left({x_1^0/x_2^0}\right)$ ~~and $\tau=q^2/S
=x_1^0 x_2^0$,
$q$ being the momentum of the dilepton pair in the DY process
and of the Higgs boson in the Higgs boson production,
$S=(p_1+p_2)^2$, 
where $p_i$ are the momenta of
incoming hadrons $P_i~(i=1,2)$.
For the DY process, $I=q$ and $\sigma^I=d\sigma^{q}(\tau,q^2,y)/dq^2$
with $q^2$ the invariant mass of the final state dilepton pair,
i.e, $q^2=m^2_{l^+l^-}$ and for the Higgs boson production through gluon
fusion, $I=g$ and $\sigma^I=\sigma^g(\tau,q^2,y)$.  The function $W^I$
can be expressed in terms of the parton distribution functions  
$f_a(x_1,\mu_F^2)$ and $f_b(x_2,\mu_F^2)$ renormalized at the
factorization scale $\mu_F$,
\begin{eqnarray}\label{wi}
W^I &=&\sum_{ab=q,\overline q,g}
\int_{x_1^0}^1 {dz_1 \over z_1}\int_{x_2^0}^1 {dz_2 \over z_2}~ {\cal H}^I_{ab}\left({x_1^0 \over z_1},{x_2^0\over z_2},
\mu_F^2\right)
\nonumber \\
&&
\times
\Delta^I_{d,ab} (z_1,z_2,q^2,\mu_F^2,\mu_R^2) \,,
\end{eqnarray}
with
\begin{align}
{\cal H}^q_{ab}(x_1,x_2,\mu_F^2)&=
f^{P_1}_a(x_1,\mu_F^2)~ f^{P_2}_b(x_2,\mu_F^2)\,,
\nonumber\\[2ex]
{\cal H}^g_{ab}(x_1,x_2,\mu_F^2)&=
x_1~ f^{P_1}_a(x_1,\mu_F^2)~ x_2~ f^{P_2}_b(x_2,\mu_F^2)\,, \,
\end{align}
where $x_i~(i=1,2)$ are the momentum fractions of the partons in the
incoming hadrons.
The threshold contribution to the rapidity distribution denoted by
$\Delta^{\rm SV}_{d,I}(z_1,z_2,q^2,\mu_R^2,\mu_F^2)$
is found to be
\begin{align}\label{delta}
\Delta^{\rm SV}_{d,I} ={\cal C} \exp
\Big({\Psi^I_d(q^2,\mu_R^2,\mu_F^2,\zo,\zt,\epsilon)}\Big)\, \Big|_{\epsilon = 0} \,,
\end{align}
where $\Psi^I_d $ are finite distributions
computed in $4+\epsilon$ space-time dimensions with $\zo = 1- z_1$ and $\zt = 1-z_2$:
\begin{align} \label{psi}
\Psi^I_d &=
\Big(
\ln \Big(Z^I(\hat a_s,\mu_R^2,\mu^2,\epsilon)\Big)^2
\nonumber\\
& 
+ \ln \big|\hat F^I(\hat a_s,Q^2,\mu^2,\epsilon)\big|^2
\Big)
\delta(\zo) \delta(\zt)
\nonumber\\
& 
+ 2~ \Phi^{~I}_d(\hat a_s,q^2,\mu^2,\zo,\zt,\epsilon)
\nonumber\\
& 
- {\cal C} \Big( \ln \Gamma_{II}(\hat a_s,\mu^2,\mu_F^2,\zo,\epsilon)~ \delta(\zt) + (\zo \leftrightarrow \zt)    \Big) 
\end{align}
The definition of double Mellin convolution 
${\cal C}$ can be found in \cite{Ravindran:2006bu}.
We drop all the regular functions that result from
various convolutions.
The bare form factors are denoted by $\hat F^I$
with $Q^2=-q^2$.
The overall operator renormalization constant
for the DY process, $Z^q = 1$ and for the Higgs boson,
$Z^g $ is known up to the three loop level \cite{renorm} in QCD. 
$\Phi^{~I}_d $ are
called the soft distribution functions and
$\Gamma_{II}$ are the mass factorization kernels.
$\mu$ is the scale introduced to define the dimensionless strong coupling 
constant $\hat a_s=\hat g_s^2/16 \pi^2$ in dimensional regularization
and $a_s(\mu_R^2)$ is the renormalized strong coupling constant 
which is related to $\hat a_s$ through the 
renormalization constant $Z (a_s(\mu_R^2))$, i.e., $\hat a_s  = (\mu/\mu_R)^\epsilon 
Z (\mu_R^2) S_\epsilon^{-1} a_s(\mu_R^2)$, $S_\epsilon = \exp[(\gamma_E-\ln 4 \pi) \epsilon/2]$.
The fact that $\Delta^{\rm SV}_{d,I}$ are finite in the limit
$\epsilon \rightarrow 0$ implies
that the pole structure of the soft distribution functions should be
similar to that of $\hat F^I$ and $\Gamma_{II}$.
We find that they must satisfy Sudakov type differential equations which
the form factors $\hat F^I$ also satisfy:
\begin{align*}
&q^2 {d \over dq^2}\Phi^{~I}_d =
{1 \over 2 }
\Big[\overline{K}^{I}_d  + \overline{G}^{I}_d  
\Big]\,,
\end{align*}
where the constants $\overline{K}^{I}_d  (\hat a_s,{\mu_R^2 \over \mu^2},\zo,\zt,\epsilon )$ are proportional to  the 
singular terms in $\epsilon$ and the $\overline G^{I}_d  (\hat a_s,{q^2 \over \mu_R^2}, {\mu_R^2 \over \mu^2},\zo,\zt,\epsilon )$
are finite functions of $\epsilon$.
It is straightforward to solve the above
differential equations yielding
\begin{align}
&\Phi^{~I}_d =
\sum_{i=1}^\infty \hat a_s^i \left({q^2 \zo \zt
\over \mu^2}\right)^{i {\epsilon \over 2}}\!\! S_{\epsilon}^i
% \nonumber\\
\left({(i~\epsilon)^2 \over 4 \zo \zt} \right)
\hat \phi^{~I,(i)}_d(\epsilon)\,,
\end{align}
where
\begin{eqnarray}
\hat \phi^{~I,(i)}_d(\epsilon)=
{1 \over i \epsilon} \Big[ \overline K^{~I,(i)}_d(\epsilon)
+ \overline {G}^{~I,(i)}_d(\epsilon)\Big]\,.
\end{eqnarray}
The constants $\overline K^{~I,(i)}_d(\epsilon)$
are determined 
by expanding $\overline K^I_d$ in powers of the
bare coupling constant $\hat a_s$, i.e.,
\begin{align}
&\overline K^I_d 
% \nonumber\\
=\delta(\zo)\delta(\zt) \sum_{i=1}^\infty \hat a_s^i
\left({\mu_R^2 \over \mu^2}\right)^{i {\epsilon \over 2}}S^i_{\epsilon}~
\overline K^{~I,(i)}_d(\epsilon) \,,
\end{align}
and solving the RG equation for
$\overline K^I_d $.
We find that $\overline K^{~I,(i)}_d(\epsilon)$ are identical to $\overline K^{~I,(i)}(\epsilon)$ given
in \cite{Ravindran:2006cg}.
The constants $\overline {G}^{~I,(i)}_d(\epsilon)$ are related to the finite
boundary functions $\overline G^I_d( \hat{a}_s, 1, {q^2 \over \mu^2}, \zo, \zt,\epsilon)$.
Defining the $\overline {\cal G}_{d,i}^I(\epsilon)$ through the relation
\begin{align*}
& \sum_{i=1}^\infty \hat a_s^i
\left( {q^2 \zo \zt \over \mu^2}\right)^{i{\epsilon \over 2}}
S^i_{\epsilon}
\overline G_d^{~I,(i)}(\epsilon)
 =
\sum_{i=1}^\infty a_s^i\left(q^2 \zo \zt \right)
\overline {\cal G}^{~I}_{d,i}(\epsilon) 
\end{align*}
and demanding the finiteness of $\Delta^{\rm SV}_{d,I}$ given in
Eq.(\ref{delta}),
we find that the structure of $\overline {\cal G}^{~I}_{d,i}(\epsilon)$ is similar to that of the corresponding $G^{~I} (\epsilon)$ in the form factors \cite{Ravindran:2006cg}, that is
\begin{equation}
 \overline{{\cal G}}^I_{d,i} (\epsilon) = - f^I_i + \overline{C}^I_i + \sum_{k=1}^{\infty} \epsilon^k \overline{{\cal G}}_{d,i}^{I,k}
\end{equation}
where $\C^I_1 = 0$, $\C^I_2 = - 2 \beta_0 \g_{d,1}^{I,1}$ ,
$\C^I_3 = - 2 \beta_1 \g_{d,1}^{I,1} - 2 \beta_0 ( \g_{d,2}^{I,1} + 2 \beta_0  \g_{d,1}^{I,2})$, 
$f_i^I$ are given in~\cite{3lffmoch}
and $\beta_i$ are the coefficients of the QCD $\beta$ function of $a_s(\mu_R^2)$,
$\mu_R^2 d a_s(\mu_R^2)/d\mu_R^2 = \epsilon a_s (\mu_R^2) / 2  -\sum_{i=0}^\infty \beta_i a_s^{i+2}(\mu_R^2)$.
%%%%%%%%%%%%%%%%%%%%%%%%%%%%%%%%%%%%%%%
The constants $\overline {\cal G}^{I,k}_{d,i} $ can be expressed in terms of 
$\overline {\cal G}^{I,k}_{i} $ using the following relation
\begin{align}
\int_0^1 
dx_1^0 \int_0^1 
dx_2^0 \left(x_1^0 x_2^0\right)^{N-1}
{d \sigma^I \over d Y}
=\int_0^1 d\tau~ \tau^{N-1} ~\sigma^I\,,
\label{iden}
\end{align}
where the $\sigma^I$ are now known for both DY and the Higgs boson production up to the N$^3$LO level in the threshold limit
\cite{Anastasiou:2014vaa, Ahmed:2014cla, Li:2014bfa}.
In the threshold limit, $N\rightarrow \infty$,  we find the following relation exact
to all orders in $\epsilon$,
\begin{eqnarray}
\hat \phi^{I,(i)}_d(\ep)=
{\Gamma(1+i~\ep) \over \Gamma^2\left(1+i{\ep \over 2}\right)}
\hat \phi^{I,(i)}(\ep).
\end{eqnarray}
where $\hat \phi^{I,(i)}(\ep)$ can be found in \cite{Ravindran:2006bu}.
Substituting ${Z^I}$, ${\hat F}^I$, and ${\Phi}^I_{d}$ and $\Gamma_{II}$ in Eq.(\ref{psi}),
and using eqn.(\ref{delta}), 
we obtain $\Delta_{d,I}^{\rm SV}$ in powers
of $a_s(\mu_R^2)$ as
\begin{align}
&\Delta_{d,I}^{\rm SV}(z_1, z_2) = 
\sum_{i=0}^\infty a_s^i(\mu_R^2) \Delta_{d,I,i}^{\rm SV} (z_1, z_2, \mu_R^2) \,, ~~~~ \text{where}
\nonumber\\
&
\Delta_{d,I,i}^{\rm SV} =
\Delta_{d,I,i}^{\rm SV} |_{\delta \delta}
\delta(\zo) \delta(\zt) 
+ \sum_{j=0}^{2i-1} 
\Delta_{d,I,i}^{\rm SV} |_{\delta {\cal D}_j}
\delta(\zt) {\cal D}_j 
\nonumber\\ 
& 
+ \sum_{j=0}^{2i-1} 
\Delta_{d,I,i}^{\rm SV} |_{\delta \overline{{\cal D}}_j}
\delta(\zo) \overline{{\cal D}}_j 
+ \sum_{j \circledS k}
\Delta_{d,I,i}^{\rm SV} |_{{\cal D}_j \overline{{\cal D}}_k}
 {\cal D}_j  \overline{{\cal D}}_k \, , \label{delsv}
\end{align}
\begin{align}
\text{with}~
{\cal D}_i=\Bigg[{\ln^i(1-z_1) \over (1-z_1)}\Bigg]_+ \,,
\bar{{\cal D}}_i=\Bigg[{\ln^i(1-z_2) \over (1-z_2)}\Bigg]_+  .
\end{align}
The symbol $j \circledS k$ implies $j, k \geq 0$ and $ j + k \leq (2 i - 2)$.
Terms proportional to ${\cal D}$ and/or $\overline {\cal D}$ in Eq.(\ref{delsv}) 
were obtained in \cite{Ravindran:2006bu} and the first term is possible to calculate 
as the results for the threshold N$^3$LO QCD corrections are now available for
DY \cite{Ahmed:2014cla} and the Higgs boson \cite{Anastasiou:2014vaa}
productions.  
\begin{widetext}
Setting $\mu_R^2=\mu_F^2=q^2$, in the following, we present this contribution along with  
the constants $\overline {\cal G}^{I,k}_{d,i}$ that are needed to determine the
soft distribution function $\Phi^I_d$ up to N$^3$LO level using $C_I=C_F, C_A$ for 
$I=q, g$, respectively.
\begin{align}
&\overline{\cal G}^{~I,1}_{d,1}
= C_I~ \Big(- \zeta_2\Big) , \hspace{1cm}
\overline{\cal G}^{~I,2}_{d,1}
 = C_I~ \Big({1 \over 3} \;\zeta_3 \Big), \hspace{1cm}
\overline{\cal G}^{~I,3}_{d,1}
= C_I~ \Big({1 \over 80} \;\zeta_2^2\Big)\,,
\nonumber\\
&\overline{\cal G}^{~I,1}_{d,2}
= C_I C_A~ \Big({2428 \over 81} -{67 \over 3} \;\zeta_2
              -4 \;\zeta_2^2 -{44 \over 3} \;\zeta_3\Big)
             +C_I n_f~ \Big(-{328 \over 81} + {10 \over 3} \;\zeta_2
                +{8 \over 3} \;\zeta_3 \Big)\,,
\nonumber\\
& \overline{\cal G}^{~I,2}_{d,2} = 
C_I C_A \Big(
-\frac{319}{120} \;{\zeta_2}^2 - \frac{71}{3} \;{\zeta_2} {\zeta_3}
+\frac{202}{9} \;{\zeta_2} + \frac{469}{27} \;{\zeta_3} 
+43 \;{\zeta_5}-\frac{7288}{243}
\Big)
+ C_I n_f \Big(
\frac{29}{60} \;{\zeta_2}^2 - \frac{28}{9} \;{\zeta_2}
-\frac{70}{27} \;{\zeta_3} +\frac{976}{243}
\Big)
\nonumber\\
& \overline{\cal G}^{~I,1}_{d,3} = 
C_I C_A^2 \Big(  
\frac{17392}{315} \;{\zeta_2}^3 + \frac{1538}{45} \;{\zeta_2}^2
+\frac{4136}{9} \;{\zeta_2} {\zeta_3} 
- \frac{379417}{486} \;{\zeta_2} 
+\frac{536}{3} \;{\zeta_3}^2 - 936 \;{\zeta_3} 
-\frac{1430}{3} \;{\zeta_5} +\frac{7135981}{8748}
\Big)
\nonumber\\
&+ C_I C_A n_f \Big(
-\frac{1372}{45} \;{\zeta_2}^2 - \frac{392}{9} \;{\zeta_2} {\zeta_3}
+\frac{51053}{243} \;{\zeta_2} 
+ \frac{12356}{81} \;{\zeta_3} 
+\frac{148}{3} \;{\zeta_5} - \frac{716509}{4374}
\Big)
+ C_I C_F n_f \Big(
\frac{152}{15} \;{\zeta_2}^2 - 40 \;{\zeta_2} {\zeta_3} 
\nonumber\\
&+ \frac{275}{6} \;{\zeta_2} 
+ \frac{1672}{27} \;{\zeta_3} 
+\frac{112}{3} \;{\zeta_5} - \frac{42727}{324}
\Big)
+ C_I n_f^2 \Big(
\frac{152}{45} \;{\zeta_2}^2 - \frac{316}{27} \;{\zeta_2} 
-\frac{320}{81} \;{\zeta_3} + \frac{11584}{2187}
\Big)
%%%%%%%%%%%%%%%%%%%%%%%%%%%%%%%%%%%%%%%%%%%%%
\end{align}
With $C_A = N$, $C_F = (N^2-1)/2N$, $n_f =$ no. of flavors and $n_{f,v}$ given in~\cite{Baikov:2009bg}, the $\delta \delta$ parts of Eq.(\ref{delsv}) for $I=q,g$ are  
\begin{align}
%%%%%%%%%%%%%%%%%%%%%%%%%%%%%%%%%%%%%%%%%%%%%%%%%%%%%%%%%%% DY delta part
& \Delta_{d,q,3}^{\rm SV}|_{\delta \delta} =
{C_A}^2 {C_F} \Big(\frac{24352}{315} \;{\zeta_2}^3 
- \frac{2921}{135} \;{\zeta_2}^2 - 588 \;{\zeta_2} {\zeta_3}
+\frac{99289}{81} \;{\zeta_2} - \frac{400}{3} \;{\zeta_3}^2 
+\frac{125105}{81} \;{\zeta_3} - 204 \;{\zeta_5}-\frac{1505881}{972}\Big)
\nonumber\\
&
+{C_A} {C_F}^2 \Big(-\frac{78272}{315} \;{\zeta_2}^3 
+ \frac{137968}{135} \;{\zeta_2}^2 + \frac{10736}{9} \;{\zeta_2} {\zeta_3}
-\frac{39865}{27} \;{\zeta_2} + \frac{1264}{3} \;{\zeta_3}^2 
- \frac{5972}{3} \;{\zeta_3} - \frac{7624}{9} \;{\zeta_5}
+\frac{74321}{36}\Big)
\nonumber\\
&
+{C_A} {C_F} {n_f} 
\Big(-\frac{2828}{135} \;{\zeta_2}^2 + \frac{272}{3} \;{\zeta_2} {\zeta_3}
-\frac{12112}{27} \;{\zeta_2} - \frac{19888}{81} \;{\zeta_3} - 8 \;{\zeta_5}+\frac{110651}{243}\Big)
+{C_F}^3 \Big(\frac{90016}{315} \;{\zeta_2}^3 - \frac{3164}{5} \;{\zeta_2}^2
\nonumber\\
&
-160 \;{\zeta_2} {\zeta_3} 
+ \frac{1403}{3} \;{\zeta_2} 
+\frac{736}{3} \;{\zeta_3}^2 - 460 \;{\zeta_3}+1328 \;{\zeta_5}-\frac{5599}{6}\Big)
+{C_F}^2 {n_f} \Big(-\frac{19408}{135} \;{\zeta_2}^2 - \frac{1472}{9} \;{\zeta_2} {\zeta_3}
+\frac{5848}{27} \;{\zeta_2} 
\nonumber\\
&
+ 360 \;{\zeta_3} 
- \frac{224}{9} \;{\zeta_5} 
-\frac{421}{3}\Big)
+{C_F} {n_f}^2 \Big(\frac{592}{135} \;{\zeta_2}^2 + \frac{2816}{81} \;{\zeta_2}
-\frac{304}{81} \;{\zeta_3} - \frac{7081}{243}\Big)
+ C_F \Big( \frac{N^2 - 4}{N} \Big) n_{f,v} \Big(
-\frac{4}{5} \;{\zeta_2}^2 
\nonumber\\
&
+ 20 \;{\zeta_2}+\frac{28}{3} \;{\zeta_3}-\frac{160}{3} \;{\zeta_5}+8
\Big)
%%%%%%%%%%%%%%%%%%%%%%%%%%%%%%%%%%%%%%%%%%%%%%%%%%%%%%%%%%%%%%%%%%%%%%%%%
\\[1ex] 
%%%%%%%%%%%%%%%%%%% Higgs delta parts
& \Delta_{d,g,3}^{\rm SV}|_{\delta \delta} = 
{C_A}^3 \Big(\frac{12032}{105} \;{\zeta_2}^3 
+ \frac{40432}{135} \;{\zeta_2}^2 - 88 {\zeta_2} \;{\zeta_3}
+\frac{41914}{27} \;{\zeta_2} + \frac{1600}{3} \;{\zeta_3}^2
- \frac{54820}{27} \;{\zeta_3} + \frac{1364}{9} \;{\zeta_5}
+\frac{215131}{81}\Big)
\nonumber\\
&
+{C_A}^2 {n_f} 
\Big(\frac{1240}{27} \;{\zeta_2}^2 - 272 \;{\zeta_2} {\zeta_3}
-\frac{7108}{27} \;{\zeta_2} + \frac{2536}{27} \;{\zeta_3}
+\frac{1192}{9} \;{\zeta_5} - \frac{98059}{81}\Big)
+{C_A} {C_F} {n_f} 
\Big(\frac{176}{45} \;{\zeta_2}^2 + 288 \;{\zeta_2} {\zeta_3}
\nonumber\\
&
-\frac{2270}{9} \;{\zeta_2} 
+ 400 \;{\zeta_3}
+160 \;{\zeta_5} - \frac{63991}{81}\Big)
+{C_A} {n_f}^2 \Big(-\frac{208}{15} \;{\zeta_2}^2
-\frac{64}{3} \;{\zeta_2} + \frac{112}{3} \;{\zeta_3}+\frac{2515}{27}\Big)
+{C_F}^2 {n_f} 
\Big(\frac{592}{3} \;{\zeta_3} 
\nonumber\\
&
- 320 \;{\zeta_5}+\frac{608}{9}\Big)
+{C_F} {n_f}^2 \Big(-\frac{32}{45} \;{\zeta_2}^2 - \frac{184}{9} \;{\zeta_2} 
-\frac{224}{3} \;{\zeta_3} + \frac{8962}{81}\Big)
\end{align}
\begin{center}
\begin{table}[h!]
\begin{tabular}{ c  c  c  c  c  c  c  c  c  c  c  c  c  c  c  c  c }
    \hline\hline
    &
    $~\delta \delta~$ &
    $~\delta \bar{\D}_0~$ & $~\delta \bar{\D}_1~$ & $~\delta \bar{\D}_2~$ & 
    $~\delta \bar{\D}_3~$ & $~\delta \bar{\D}_4~$ & $~\delta \bar{\D}_5~$ &
    $~\D_0 \bar{\D}_0~$ & $~\D_0 \bar{\D}_1~$ & 
    $~\D_0 \bar{\D}_2~$ & $~\D_0 \bar{\D}_3~$ & $~\D_0 \bar{\D}_4~$ & 
    $~\D_1 \bar{\D}_1~$ & $~\D_1 \bar{\D}_2~$ & $~\D_1 \bar{\D}_3~$ &
    $~\D_2 \bar{\D}_2~$ \\      
    \hline
    \% &
    73.3 & 16.0 & 9.1 & 31.4 & 1.0 & -9.9 & -23.1 &
    -13.7 & -10.7 & -0.3 & 3.1 & 7.3 & -0.2 & 3.8 & 8.6 & 4.2 \\
    \hline\hline
  \end{tabular}
 \caption{Relative contributions of pure N$^3$LO terms.}
 \label{table:perc}
\end{table}
\begin{table}[h!]
\begin{tabular}{ l  c  c  c  c  c  c  c  c  c  c }
    \hline\hline
    $Y$ & ~~~~~0.0~~~~ & ~~~~~0.4~~~~ & ~~~~~0.8~~~~ & ~~~~~1.2~~~~ & ~~~~~1.6~~~~ & ~~~~~2.0~~~~ & ~~~~~2.4~~~~ & ~~~~~2.8~~~~ & ~~~~~3.2~~~~ & ~~~~~3.6~~~~ \\
    \hline
    NNLO & 11.21 & 10.96 & 10.70 & 9.13 & 7.80 & 6.10 & 4.23 & 2.66 & 1.40 & 0.54 \\
%    \hline
    NNLO$_{\rm SV}$  & 9.81 & 9.61 & 8.99 & 8.00 & 6.71 & 5.21 & 3.66 & 2.25 & 1.14 & 0.42 \\
%    \hline
    NNLO$_{\rm SV}$(A)  &  10.67 & 10.46 & 9.84 & 8.82 & 7.48 & 5.90 & 4.24 & 2.69 & 1.42 & 0.56 \\
%    \hline
    N$^3$LO$_{\rm SV}$  & 11.62 & 11.36 & 11.07 & 9.44 & 8.04 & 6.27 & 4.33 & 2.70 & 1.40 & 0.53 \\
%    \hline
    N$^3$LO$_{\rm SV}$(A) & 11.88 & 11.62 & 11.33 & 9.70 & 8.30 & 6.51 & 4.54 & 2.88 & 1.53 & 0.60 \\
%    \hline
    $K$3 & 2.31 & 2.29 & 2.36 & 2.21 & 2.17 & 2.07 & 1.89 & 1.70 & 1.63 & 1.51 \\
    \hline\hline
 \end{tabular}
 \caption{Contributions of exact NNLO, NNLO$_{\rm SV}$, N$^3$LO$_{\rm SV}$, and $K3$.}
 \label{table:nnlo}
\end{table}

\end{center}
\end{widetext}
We present the relative contributions in percentage
of the pure N$^3$LO terms in Eq.(\ref{delsv}) 
with respect to $\Delta^{\rm SV}_{d,g,3}$, for rapidity $Y$ = 0 in Table~\ref{table:perc}.
The notation $\D_i \bar{\D}_j$ corresponds to the sum of the contributions coming from 
$\D_i \bar{\D}_j$ and $\D_j \bar{\D}_i$.
We have used $\sqrt{s} = 14$ TeV for the LHC, $G_F = 4541.68$ pb, the $Z$ boson 
mass $m_Z$ = 91.1876 GeV, 
top quark mass $m_t$ = 173.4 GeV
and the Higgs boson mass $m_H$ = 125.5 GeV throughout. 
For the Higgs boson production, we use the effective theory where top quark is integrated out in the large $m_t$ limit.
The strong coupling constant $\alpha_s (\mu_R^2)$ is evolved 
using the 4-loop RG equations with 
$\alpha_s^{\text{N$^3$LO}} (m_Z ) = 0.117$ and for parton density sets we use 
MSTW 2008NNLO \cite{Martin:2009iq}, as N$^3$LO evolution kernels are not yet available. 
In \cite{Forte:2013mda}, Forte \textit{et al.} pointed out that the Higgs boson cross sections will remain unaffected with this shortcoming. However, for the DY process, it is not clear whether the same will be true.
We find that the contribution from the $\delta(\zo) \delta (\zt)$ part is the largest.
The dependence on the renormalization and factorization scales can by studied by
varying them in the range $\frac{m_H}{2} <\mu_R,\mu_F<2 m_H$. We find that the inclusion of the 
threshold correction at N$^3$LO further reduces their dependence.  
For the inclusive Higgs boson production, we find that
about 50\% of exact NNLO contribution comes from   
threshold NLO and NNLO terms.  It increases to 80\% if we use exact NLO and threshold NNLO terms.
Hence, it is expected that the rapidity distribution of the Higgs boson
will receive a significant contribution from the threshold region compared to
inclusive rate due to the soft emission over the entire range of $Y$.
Our numerical study with threshold enhanced NNLO rapidity distribution confirms our expectation. 
Comparing our threshold NNLO results
against exact NNLO distribution using the FEHiP \cite{Anastasiou:2005qj} code , 
we find that about $90\%$ of
exact NNLO distribution comes from the threshold region as can be seen from Table~\ref{table:nnlo}, in accordance with~\cite{Becher:2007ty}, where it was shown that for low $\tau~(m_H^2/s \approx 10^{-5})$ values the threshold terms are dominant, thanks to the inherent property of the matrix element, which receives the largest radiative corrections from the phase-space points corresponding to Born kinematics.  
Here we have used the exact results up to the NLO level. 
Because of an inherent ambiguity in the definition of 
the partonic cross section at threshold one can multiply a factor $z g(z)$, where $z=\tau/x_1 x_2$ and $\lim_{z \rightarrow 1} g(z) = 1$, with the partonic flux and divide the same in the partonic cross section for an inclusive rate. In~\cite{Catani:2003zt,Kramer:1996iq} this was exploited to take into account the subleading collinear logs also, thereby making the threshold approximation a better one. Recently, Anastasiou \textit{et al.} used this in~\cite{Anastasiou:2014vaa} to modify the partonic flux keeping the partonic cross section unaltered to improve the threshold effects.
Following \cite{Anastasiou:2014vaa,Herzog:2014wja}, we  
introduce
$G (z_1,z_2)$ such that $\lim_{z_1,z_2 \rightarrow 1} G = 1$ in 
(\ref{wi}): 
\begin{eqnarray}
W^I &=&\sum_{ab=q,\overline q,g}
\int_{x_1^0}^1 {dz_1 \over z_1}\int_{x_2^0}^1 {dz_2 \over z_2}~ {\cal H}^I_{ab} ~ G(z_1,z_2)
\nonumber \\
&&\times~
\lim_{z_1,z_2 \rightarrow 1} \Big[ \frac{ \Delta^I_{d,ab} (z_1,z_2)}{G(z_1,z_2)} \Big] \,.
\end{eqnarray}
We also find that with the choice $G(z_1,z_2)=z_1^2 z_2^2$, the threshold NNLO results  are
remarkably close to the exact ones for the entire range of $Y$ [see Table~\ref{table:nnlo}, denoted by $(A)$].  
This clearly demonstrates the dominance
of threshold contributions to rapidity distribution of the Higgs boson production at the NNLO level.  
Assuming that the trend will not change drastically beyond NNLO, we present numerical values for
N$^3$LO distributions for $G(z_1,z_2)=1, z_1^2 z_2^2$, respectively, as N$^3$LO$_{\rm SV}$ and N$^3$LO$_{\rm SV}(A)$
in Table~\ref{table:nnlo}. 
The threshold N$^3$LO terms give $6 \% (Y = 0)$ to $12 \% (Y = 3.6)$ additional 
correction over the NNLO contribution to the inclusive Higgs production.
Finally, in Table \ref{table:nnlo}, we have presented $K3 =$ N$^3$LO$_{\rm SV}$/LO as a function of $Y$ in order to demonstrate 
the sensitivity of higher order effects to the rapidity $Y$. 
% 

%%%%%%%%%%%%%%%%%%%%%%%%%%%%%%%%%%%%%%%
%
To summarize, we present full threshold enhanced N$^3$LO QCD corrections to 
rapidity distributions of the dilepton pair in the DY process and of the Higgs boson in gluon-gluon fusion at the LHC.
We show that the infrared structure of QCD amplitudes, 
in particular, their factorization properties,  
along with Sudakov resummation of soft gluons and renormalization group invariance provide an elegant
framework to compute these threshold corrections systematically for rapidity distributions 
order by order in QCD perturbation theory.   
The recent N$^3$LO results for inclusive DY and Higgs boson production cross sections
at the threshold provide crucial ingredients to obtain $\delta(\overline z_1) \delta(\overline z_2)$
contribution of their rapidity distributions for the first time.  
We find that this contribution numerically dominates over the rest of the terms in $\Delta^{\rm SV}_{d,g,3}$ at
the LHC. Inclusion of N$^3$LO contributions reduces the scale dependence further.  
We also demonstrate the dominance of the threshold contribution to rapidity distributions by comparing it against the exact NNLO for two different choices of $G(z_1,z_2)$. Finally, we find that threshold N$^3$LO rapidity distribution with $G(z_1,z_2)=1,z_1^2 z_2^2$ shows a moderate effect over NNLO distribution.

%{\bf Acknowledgements:} 
We thank F. Petriello for providing the FEHiP code and fruitful discussions.
T.A., M.K.M. and N.R. thank IMSc for
providing hospitality. 
We thank M. Mahakhud for discussion.
The work of T.A., M.K.M. and N.R. has been partially supported by funding from RECAPP,
DAE, Govt. of India.

\vspace{-1.2cm}

\theendnotes

\vspace{0.2cm}

%%%%%%%%%%%%%%%%%%%%%%%%%%%%%%%%%%%%%%%%

\end{document}